\documentstyle[12pt]{article}
\begin{document}

\begin{center}
\vspace*{5cm}
{\large\bf
An antiproton-proton partial-wave analysis} \\
\vspace{1cm}
R. Timmermans \\
{\it Theory Division, Los Alamos National Laboratory,
     Los Alamos, NM 87545, USA} \\
\vspace{.5cm}
Th.A. Rijken and J.J. de Swart \\
{\it Institute for Theoretical Physics,
     University of Nijmegen, Nijmegen, The Netherlands}
\end{center}
\vspace{.2cm}

\begin{center}
{\bf I. Introduction}
\end{center}
Partial-wave analyses (PWAs) have a long history
in the fields of $\pi N$ and $N\!N$ scattering.
Due to the poor quality of low-energy antiproton beams
and the resulting absence of accurate experimental data,
analogous model-independent studies of the much more complex
$\overline{p}p$ system have in the past always been impossible.
In recent years, however, experimental progress has been
very significant, in particular due to the coming in 1983
of the Low-Energy Antiproton Ring (LEAR) facility at CERN.
While in the pre-LEAR era spin-dependent observables and
charge-exchange ($\overline{p}p \rightarrow \overline{n}n$)
data were almost nonexistent, the situation between 400 and
925 MeV/c is now quite good: the LEAR collaborations PS172,
PS173, PS198, and PS199 have measured a variety of observables
with impressive accuracy. High-quality analyzing-power
data have been obtained for the elastic~\cite{Kun88}
and charge-exchange~\cite{Bir90} reactions. Very recently,
even charge-exchange depolarization data have become
available~\cite{Bir93}. Unfortunately, the practical difficulties
involved in constructing a high-quality ``cooled'' antiproton beam
of lower momentum are large. Consequently, the $\overline{p}p$
database below about 400 MeV/c is still by far not as good as
one would like, in striking contrast to the $pp$ case where very
accurate data exist as low as $T_L=0.35$ MeV ($p_L=25$ MeV/c).
It also remains an outstanding experimental challenge to
construct a polarized antiproton beam to further probe the spin
structure of the interaction.
 
During the last 10 years a new method has been developed by
the Nijmegen group to perform PWAs of the abundant and accurate
$N\!N$ ($pp$ and $np$) scattering data below $T_L=350$
MeV~\cite{Ber90,Klo93}.
With the now available high-quality data from LEAR and KEK,
we have been able to extend the methods used in these $N\!N$
PWAs to perform an energy-dependent PWA of all
$\overline{p}p$ scattering data below $p_L=925$ MeV/c
($T_L=379$ MeV). This work was started in 1987~\cite{Tim91}
and has only recently been finished~\cite{Tim94}.
The same methods of PWA have also been
applied~\cite{Tim92} to the strangeness-exchange reaction
$\overline{p}p \rightarrow \overline{\Lambda}\Lambda$,
for which the PS185 group at LEAR has obtained beautiful data.
In the next section we review the theoretical ideas behind
these Nijmegen PWAs, and in section III we apply these
ideas and methods to the case of $\overline{p}p$ scattering.
In section IV some results of this $\overline{p}p$ PWA are
presented and discussed.

After almost a decade of LEAR, it is fair to say
that in this field theory has some catching up to
do with respect to experiment. Since the partial-wave
amplitudes or the phase-shift parameters are in a sense
the meeting ground between theory and experiment, the
results of the present PWA should be very useful in many
ways. They can be used to improve models~\cite{Cot82,Tim84}
for the $\overline{N}\!N$ interaction. Apart from the fact that
this provides independent and complementary~\cite{Tim94} information
about the spin- and isospin structure of the $N\!N$ force,
the $\overline{N}\!N$ interaction is needed as input in many
other $\overline{p}p$ subfields. Studies of for instance
protonium (the $\overline{p}p$ atom)
or specific annihilation processes like
$\overline{p}p \rightarrow \pi^+\pi^-$, $K^+K^-$
require a realistic treatment of the initial $\overline{p}p$
interaction. At the same time, this PWA could be helpful
in planning new experiments at LEAR, the future of which
is of course crucial to this field.
 
\begin{center}
{\bf II. Methods of partial-wave analysis}
\end{center}
The hallmark of the Nijmegen energy-dependent PWAs is the
sophisticated manner in which the energy dependence of the
partial-wave amplitudes is parametrized. At the basis of the
PWA is the trivial observation that in the low-energy region
(long wave lengths) the long-range interaction is very important.
It is this long-range interaction that is responsible for the
rapid variations with energy of the scattering amplitudes.
Short-range interactions lead to much slower energy
variations of the amplitudes. One usually looks for
a function in the problem that one can parametrize as
easily as possible, i.e. one that does not contain the
contributions from these long-range processes.
Because these long-range interactions are at the same time
model independent (in the sense that they are or at least
should be the same in all $N\!N$ and $\overline{N}\!N$ models),
they can then be taken into account separately and exactly.

It is, of course, not a good idea to try to parametrize the
partial-wave $S$ matrix itself, since it does contain all of
these long-range effects. As a function of complex energy, the $S$
matrix has a (kinematical) right-hand unitarity cut, other right-hand
cuts due to the coupling to inelastic channels, and (dynamical)
left-hand cuts due to particle exchanges. The left-hand cuts
that are the closest to the origin $T_L=0$ correspond to the
longest-range processes. The left-hand cuts that start far away
from the origin are due to the short-range interactions. For
instance, the infinite-range Coulomb potential ($V\sim 1/r$)
produces an essential singularity and a branch point at $T_L=0$,
vacuum polarization ($V\sim\exp(-2m_{e}r)/r^{5/2}$)
produces a cut at $T_L=-0.6$ keV, and one-pion exchange
($V\sim\exp(-m_{\pi}r)/r$) leads to a cut starting at $T_L=-9.7$ MeV.
One sees that the crux is to find a quantity in which the cuts
nearest to the origin are not present. This quantity then
allows an {\it analytical} parametrization in energy or $k^2$
in an enlarged domain up to the next left- or right-hand
cut present.

A familiar example of such a quantity with improved
analyticity properties is the modified effective-range
function~\cite{Ber90,Hae82}. The Coulomb-modified effective-range
function for the $pp$ $^1S_0$ state was originally derived
(in a rather intuitive way) by Landau and Smorodinsky~\cite{Lan44}.
When only the Coulomb potential is present the boundary condition
for the radial wave function $\Phi(0)=0$ is of course satisfied
by $F$, the regular Coulomb wave function (for $\ell=0$).
Suppose that there is an additional short-range interaction.
When the wave length is very large (very low energy), one
can take the limit in which the range of this additional
(strong) interaction goes to zero. Then its presence is {\it only}
revealed by a modified boundary condition at $r=0$, which
is now satisfied by a linear combination of $F$ and $G$,
the irregular Coulomb wave function
\begin{equation}
   \Phi(r) \: = \: F(r) \cot\delta_0 + G(r) \:\: ,
\label{eq:lev1}
\end{equation}
where $\delta_0$ is the nuclear phase shift in the presence
of the Coulomb interaction ($\delta(^1S_0)=\delta_0+\sigma_0$),
as can be seen from the asymptotic behavior of $\Phi$.
An equation for $\cot\delta_0$ can then be obtained by evaluating
the logarithmic derivative of the wave function, which we call
$P(k,r)$, for $k\rightarrow 0$. In the $np$ case this quantity
$P(k,0)=k\cot\delta_0$ approaches a constant:
\begin{equation}
   P(k,\varepsilon) \: = \:
   \left(\frac{d\Phi}{dr}/\Phi\right)_{r=\varepsilon}
   \rightarrow \:\:\: -\frac{1}{a} \:\: .
\label{eq:lev2}
\end{equation}
In the $pp$ case, the evaluation has to
be done at $r=\varepsilon$ because of a term
$\ln\varepsilon$ that appears due to the singular behavior
of $G$. This term one absorbes in the constant $-1/a$,
along with some further constant terms. Then one lets
$\varepsilon\rightarrow 0$ and immediately obtains the
Coulomb-modified (``zero-range'') effective-range function.
It can be shown that after these manipulations
the resulting left-hand side of Eqn.~(\ref{eq:lev2})
is an analytical (actually meromorphic) function of the energy,
so that the right-hand side can be written as a power series
in $k^2$ (this means dropping the zero-range approximation).

The analytical expansion of the Coulomb-modified effective-range
function breaks down already at $T_L=-9.7$ MeV, where the
one-pion--exchange cut starts. It is possible to derive
a new ``pion-modified'' effective-range function from which also
this cut has been removed~\cite{Ber90}. Let the regular and
irregular wave functions for the case where only the Coulomb and
pion-exchange potentials are present be called $F_\pi$ and $G_\pi$.
(For the purpose of the present discussion, we ignore vacuum
polarization.) The wave function can then be written as
\begin{equation}
   \Phi(r) \: = \: F_\pi(r) \cot\delta_0 + G_\pi(r) \:\: ,
\label{eq:wim}
\end{equation}
where $\delta_0$ is now the phase shift due to the short-range remainder
of the strong interaction ($\delta(^1S_0)=\delta_0+\pi_0+\sigma_0$,
where $\pi_0$ is the one-pion--exchange phase shift in the
presence of the Coulomb potential). However, proceeding in similar
fashion as above, one encounters an important problem here.
The evaluation of $P(k,\varepsilon)$ has to be done numerically,
since $F_\pi$ and $G_\pi$ are not known in analytical form.
Due to the singular behavior of $G_\pi$ when $\varepsilon\rightarrow 0$,
it is very hard to maintain sufficient numerical accurary,
especially for higher orbital angular momenta.

At this point one has to realize that this numerical problem
of the modified effective-range function is really an {\it artificial}
problem: it crops up due to the singular behavior of the irregular
function of the {\it long-range} potential {\it near the origin}.
However, it is precisely this short-range interaction that
one wants to parametrize, since it is essentially unknown,
very complicated, and leads to only slow energy variations
of the scattering amplitudes. Looking at Eqn.~(\ref{eq:wim}),
one observes that it is valid for {\it any} $r$, so why not
evaluate $P(k,r)$ at a {\it finite} value $r=b$, instead of
at $r=\varepsilon$?

This is essentially what is done in the Nijmegen PWAs.
The wave functions are obtained by solving the (relativistic)
Schr\"odinger equation. Suppose one starts at a point
$r_{\infty}$ where only the Coulomb potential is present.
Integrating inwards, one picks up sequentially the
contributions (varying rapidly with energy) from the electromagnetic
potentials, one-pion exchange, and contributions (varying slower with
energy) from other meson exchanges. One then stops at a point $r=b$.
If there are no additional interactions for $r<b$, the boundary
condition $P(k,b)$ at $r=b$ is obviously satisfied by the
regular wave function corresponding to precisely this potential tail.
For small enough $b$ the model used for $r>b$ will of course
not be correct, and the boundary condition has to be modified,
as in the above examples. In practice, it works the other way:
one starts integrating at $r=b$ and $P(k,b)$ is parametrized
as a function of energy. Also the best value for $b$ is
determined by fitting the data.
In general multichannel problems $P(k,b)$ becomes a matrix.
This $P$ matrix has the required improved analyticity
properties. When there are no nearby right-hand cuts, it is an
analytical (again: actually meromorphic) function of $k^2$ in a
domain bounded by the nearest left-hand cut not removed by
including (or including incorrectly) the corresponding exchange
in the potential tail for $r>b$. It can happen, of course,
that short-range dynamics gives rise to a rapid energy variation
of the amplitudes, as in the case of a resonance. This would
have to be taken into account in the $P$ matrix,
for instance by including a pole in the parametrization.
It is seen that the formalism used in the Nijmegen PWAs
is similar to the boundary-condition approach to the
strong interactions that goes back to the work of Feshbach
and Lomon~\cite{Fes64} and earlier. The philosophy, however,
is very different. The term $P$ matrix (for ``pole''
matrix) was introduced by Jaffe and Low~\cite{Jaf79}
in the framework of the bag model.

\begin{center}
{\bf III. An antiproton-proton partial-wave analysis}
\end{center}
Let us now be more specific and apply the foregoing
ideas to the case of $\overline{p}p$ scattering.
In all the Nijmegen PWAs, the two-body scattering
process is described with the relativistic Schr\"odinger
equation~\cite{Swa78,Aus83}, which is essentially a
coordinate-space version of the Blankenbecler-Sugar equation.
It reads the same as the ordinary Schr\"odinger equation
\begin{equation}
  \left( \Delta+k^2-2mV \right) \psi({\bf r}) \: = \: 0  \:\: ,
\label{eq:psi}
\end{equation}
except that the proper relativistic relation between
energy and momentum is used. It is well known how to derive
the potentials for use in this equation~\cite{Swa78,Aus83}.
In this relativistic framework, there is no known
quantum-mechanical interpretation for
the ``wave function'' $\psi({\bf r})$. It is perhaps
best to regard it as just a tool that allows one to
compute the correct relativistic scattering amplitude
(e.g. the poles are the correct bound states). We solve
Eqn.~(\ref{eq:psi}) for the coupled $\overline{p}p$ and
$\overline{n}n$ channels. The mass difference between
proton and neutron is included in order to account for
the $\overline{n}n$ threshold at $p_L=99$ MeV/c.

The interaction in the region $r>b$ is described by  
a theoretically well-founded $\overline{N}\!N$ potential.
This potential is given by
\begin{equation}
     V\: =\: V_C + V_{M\!M} + V_{N} \:\: ,
\end{equation}
where $V_C$ and $V_{M\!M}$ are the relativistic Coulomb and
magnetic-moment interaction respectively. $V_{N}$ is the
$\overline{N}\!N$ meson-exchange potential. It consists
of one-pion exchange and the (charge-conjugated)
heavy-meson and pomeron exchanges from the 1978
Nijmegen $N\!N$ potential~\cite{Nag78}. As argued in
the previous section, the rapid energy variations of the
amplitudes due to the long-range electromagnetic interactions
and one-pion exchange are now included {\it exactly}.

Let us next turn to the parametrization of the short-range
interactions for $r<b$ by way of the $P$-matrix boundary condition
at $r=b=1.3$ fm. Due to the coupling to the annihilation channels,
the $S$ matrix has a right-hand cut starting
already to the left of $T_L=0$. (In the $pp$
case this cut starts only at the $pp \rightarrow pp\pi^0$ threshold
at $T_L=280$ MeV.) As these annihilation processes are of short range
(and so give rise to slow energy variations of the amplitudes),
this right-hand cut has to be present in the $P$ matrix, which
we therefore take to be complex. (Similarly, the effective-range
parameters for the $\overline{N}\!N$ case are complex.)
The choice of the value for $b$ is rather critical, more so than
in the $N\!N$ case (where it was taken to be $b=1.4$ fm). The best
results are obtained for $b=1.3$ fm. Since for $r>b$ we use only a
real potential, the coupling to the annihilation channels is
completely represented by the boundary condition. We conclude
therefore that the range of the annihilation process is in fact
about 1.3 fm~\cite{Tim94}.

The electromagnetic interactions that we use are adapted
from the improved Coulomb potential~\cite{Aus83}. This
potential, designed specifically for use in the relativistic
Schr\"odinger equation, contains relativistic corrections to
the static Coulomb potential and (in its off-shell behavior)
the main contributions from
the two-photon--exchange diagrams. All these effects are
included in the Nijmegen $pp$ PWA~\cite{Ber90,Klo93}, as well
as the vacuum-polarization potential. In our case it suffices
to use the following spin-dependent one-photon--exchange potentials
\begin{equation}
   V_{\gamma}(r) \: = \: -\frac{ \alpha'}{ r} \: + \:
   \frac{\mu^{2}_{p}}{4M^{2}_{p}}\:\:
   \frac{\alpha}{ r^{3}}\:S_{12}
   \: + \: \frac{8\mu_{p}-2}{ 4M^{2}_{p}}
   \frac{\alpha}{ r^{3}}\:{\bf L}\cdot{\bf S} \:\:\:\:\:
   {\rm for} \:\: \overline{p}p \rightarrow \overline{p}p \:\: ,
\end{equation}
and
\begin{equation}
   V_{\gamma}(r) \: = \: \frac{\mu^{2}_{n}}{4M^{2}_{n}}
   \frac{\alpha}{ r^{3}}\:S_{12} \:\:\:\:\:
   {\rm for} \:\: \overline{n}n \rightarrow \overline{n}n \:\: .
\end{equation}
The magnetic moments of the proton and neutron are
$\mu_p=1+\kappa_p=2.793$ and $\mu_n=\kappa_n=-1.913$,
respectively. The use of $\alpha'$ in the central potential
for $\overline{p}p \rightarrow \overline{p}p$ takes care
of the main relativistic corrections to the Coulomb potential.
It is given by $\alpha'/\alpha=2k/Mv_L$ where $v_L$ is the
velocity of the antiproton in the laboratory system. At 600
MeV/c one has for instance $v_L=0.54$ and $\alpha'/\alpha=1.135$,
a correction of $13.5\%$ to the static Coulomb potential.
The spin-orbit potential comes from the interaction of the
magnetic moment of one particle with the Coulomb field of
the other particle (and is consequently absent in
$\overline{n}n \rightarrow \overline{n}n$). It includes
a relativistic correction due to the Thomas precession.
The tensor potential comes from the interaction of the two
magnetic moments. Vacuum polarization and two-photon--exchange
effects are negligible in our case. The proper treatment of
these electromagnetic effects in the evaluation of the
scattering amplitudes is a nontrivial matter~\cite{Tim94}.
The following simple one-pion--exchange potential without
a form factor is used
\begin{equation}
 V_{\pi}(r) \: = \: f^2_{N\!N\pi} \frac{M}{\sqrt{k^2+M^2}} \:
            \frac{m^2}{m_{\pi^\pm}^2} \: \frac{1}{3}
            \left[ \mbox{\boldmath $\sigma$}_{1}\cdot
               \mbox{\boldmath $\sigma$}_{2} + S_{12}
            \left(1+\frac{3}{(mr)}+\frac{3}{(mr)^{2}} \right)\right]
            \frac{e^{-mr}}{r}  \;\; ,
\end{equation}
where $m$ is the mass of the pion and $f^2_{N\!N\pi}=0.0745$
is the rationalized pion-nucleon coupling constant~\cite{Sto93}.
The mass difference between the $\pi^0$ and
$\pi^{\pm}$ is included.

Let us finish this section with some more general remarks about PWAs.
Even for the $pp$ case, where the database is of high
quality and the observables are very well mapped out,
a PWA is impossible without a substantial amount of
theoretical input or constraints. For the $np$ and
$\overline{p}p$ PWAs, this is true a fortiori. For
instance, one has to make some assumptions about the
validity of symmetries like charge independence or
(as in our case) charge conjugation. Obviously, one
has to careful here: sometimes general physical
principles are inspired by local renormalizable field
theories and not strictly valid for extended objects like
hadrons. A good example can be found in $\pi N$ PWAs, where
one usually implements full Mandelstam analyticity~\cite{Man58}.
The amplitudes are assumed to be analytic functions of
the two complex variables $s$ and $t$ except for singularities
from the mass spectrum and unitarity. These amplitudes then
exhibit crossing symmetry. It is not clear at all to what
extent low-energy hadron dynamics actually satisfies this
symmetry.

Using strong and mostly model-independent
theoretical constraints it has turned out to be possible to
perform an {\it energy-dependent} or {\it multienergy} PWA
of the $\overline{p}p$ data. However, it is quite a different
ballgame to perform {\it energy-independent} or
{\it single-energy} $\overline{p}p$ PWAs. In a single-energy
$\overline{p}p$ PWA one has to determine in principle 20
phase-shift parameters for each $J \neq 0$ (8 for $J=0$), which
is four times as many as in a single-energy $np$ PWA~\cite{Tim94}!
Almost certainly the present $\overline{p}p$ database does not
allow satisfactory energy-independent PWAs. One has to
realize, however, that even in the $N\!N$ field the usefulness of
energy-independent PWAs is more limited than is perhaps generally
thought. When one has a {\it good} energy-dependent PWA,
the best value for a phase shift (or the pion-nucleon coupling
constant!) is definitely the one determined in the energy-dependent
PWA, and not the one from an energy-independent PWA.
One reason is that an energy-independent PWA contains no
information about the energy dependence of the amplitudes. This
makes it for instance less stable than an energy-dependent PWA 
with respect to the addition of new data to the database. Also,
a set of energy-independent PWAs is usually overparametrized compared
to a good energy-dependent PWA in the same energy region. It thus
almost certainly contains noise. For an extensive discussion of
this important point, see Ref.~\cite{Klo93}.

\begin{center}
{\bf IV. Some results of the analysis}
\end{center}
While in $N\!N$ PWAs there is essentially agreement on the
correct database (especially for $pp$), we had to spend a lot
of time and effort into collecting, scrutinizing, and
cleaning up the world set of $\overline{p}p$ scattering data,
which contains a lot of flaws and contradictory data.
Exactly the same statistical arguments were used in this
process as were used in the set-up of the Nijmegen $N\!N$
database~\cite{Ber90,Klo93}. This means for instance that
data with a very improbable high {\it or low} $\chi^2$
are rejected on statistical grounds. The resulting Nijmegen
1993 $\overline{p}p$ database in the momentum interval
119--923 MeV/c is unique in the world and consists
of $N_{\rm data}=3646$ $\overline{p}p$ data.
It is extensively discussed in Ref.~\cite{Tim94}.
In the final fit to this database
we reach $\chi^2=3801.0$ or $\chi^2/N_{\rm data}=1.04$.
The number of boundary-condition parameters needed is 30, which
is a reasonable number, in view of the fact that 21 parameters
were needed in the Nijmegen $pp$ PWA and an additional
18 in the $np$ PWA. The total number of degrees of freedom is
$N_{\rm df}=3503$, which means that $\chi^2/N_{\rm df}=1.09$.

If the database is a correct statistical ensemble and if the
theoretical model is correct, one expects that $\langle\chi^2\rangle=
N_{\rm df}=3503$ with an error of $\sqrt{2N_{\rm df}}=84$. This means
that in our PWA we are 298 or only 3.5 standard deviations away
from the expectation value for $\chi^2$. We conclude that
although there is still room for improvement, our {\it energy-dependent
solution is essentially correct statistically}. As a consequence,
the values for the phase-shift parameters (and also for the
pion-nucleon coupling constant) and the statistical errors
(obtained in the standard manner from the error matrix) are
essentially correct.

In our 1991 preliminary PWA~\cite{Tim91} we were able to
determine the charged-pion--nucleon coupling constant
$f^2_c \equiv f_{pn\pi^+}f_{np\pi^-}/2$
from the data on the charge-exchange reaction
$\overline{p}p \rightarrow \overline{n}n$, in which
only isovector mesons can be exchanged. The result
found was $f^2_c=0.0751(17)$, at the pion pole.
The error is purely statistical. In our final analysis,
we have repeated the determination of $f^2_c$,
but this time from the complete 1993 Nijmegen database.
The coupling constants of the neutral pion were kept at the
value of $f^2_{pp\pi^0}=f^2_{nn\pi^0}=0.0745$~\cite{Sto93}.
We now find $f^2_c=0.0732(11)$, at the pion pole.
This result supersedes our previous value from Ref.~\cite{Tim91}.
Again, the error is of statistical origin only. In view of
the enormous amount of work involved, it is very hard
to estimate possible systematic errors
on this result. We have checked that there are no systematic
errors due to form-factor effects or due to uncertainties in
$\rho^\pm$(770) exchange. In the Nijmegen $pp$ PWA
systematic errors could be more thoroughly investigated and
they were found to be small~\cite{Sto93}. In our case the
systematic errors are probably larger than for the $pp$ case,
but it is very encouraging that the result for $f^2_c$ is in
good agreement with recent determinations $f^2_c=0.0735(15)$
from $\pi^{\pm}p$~\cite{Arn90} scattering
and $f^2_{pp\pi^0}=0.0745(6)$ and $f^2_c=0.0748(3)$ from
$N\!N$ scattering~\cite{Sto93}. Very probably the new LEAR
experiment PS206 on $\overline{p}p \rightarrow
\overline{n}n$ will further constrain the
charged-pion--nucleon coupling constant.

\begin{figure}
\vspace*{8cm}
\includegraphics{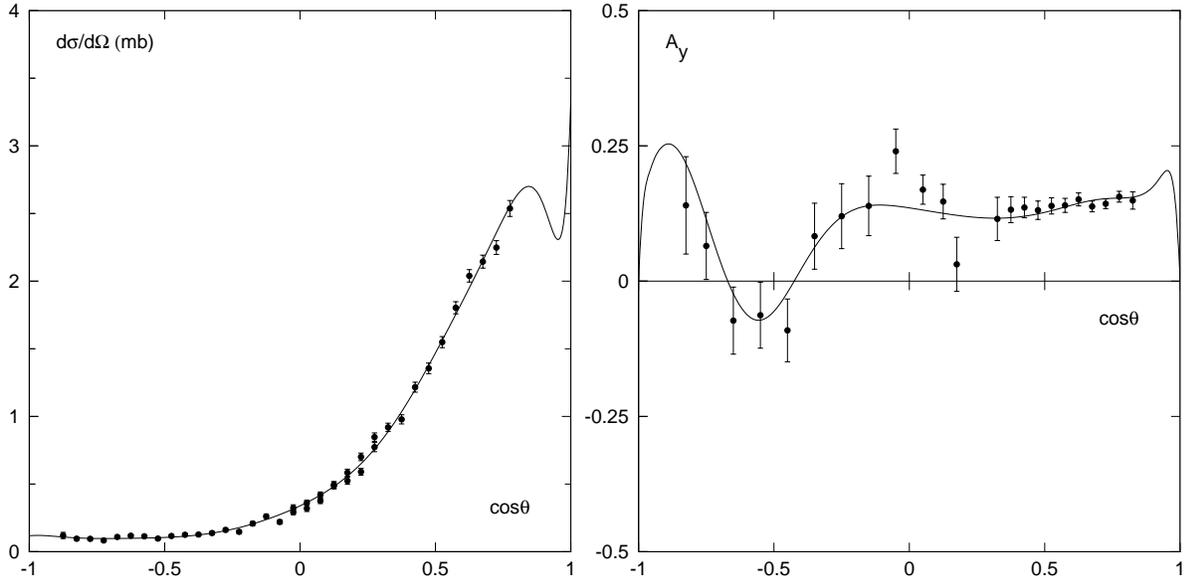}
\caption{Differential cross section at 693 MeV/c and analyzing
         power at 875 MeV/c for the charge-exchange reaction
         $\overline{p}p \rightarrow \overline{n}n$. The data are
         from PS199~\protect\cite{Bir90}. The curves are from the
         Nijmegen PWA~\protect\cite{Tim94}.}
\label{fig:ps199}
\end{figure}

In Fig.~\ref{fig:ps199} the differential cross section at 693 MeV/c
and the analyzing power at 875 MeV/c are shown for $\overline{p}p
\rightarrow \overline{n}n$. The data are from PS199~\cite{Bir90}.
One can see the truly remarkable accuracy of the cross-section data
and the analyzing-power data in the forward region. The ``dip-bump''
structure in $d\sigma/d\Omega$ at forward angles is due to the
interference of one-pion exchange and a smooth background.

The fact that the available charge-exchange data already
pin down the charged-pion coupling constant with a remarkable
small statistical error is only one example of how at present
{\it quantitative} information can be extracted from
the $\overline{p}p$ system. We can mention some subtle effects
that are also visible in the data. The use of $\alpha'$ instead
of $\alpha$, i.e. the main relativistic correction to the
static Coulomb potential, gives a drop of $\Delta\chi^2=30$,
or 5.5 standard deviations. The inclusion of the magnetic-moment
interaction gives a drop of $\Delta\chi^2=14$, or 3.7 standard
deviations. Even the use of the correct pion masses instead of an
average mass of 138 MeV is a 3 standard-deviation effect.

Since the present $\overline{p}p$ PWA is the first of its
kind, we have also proposed a convention for extracting
phase-shift and inelasticity parameters from the $S$ matrix.
In the presence of coupling to annihilation channels
the $S$ matrix describing $\overline{N}\!N$ scattering
is only a submatrix of the much larger multichannel
$S$ matrix. It is therefore still symmetric, but no longer
unitary. This doubles the number of parameters needed.
For the partial waves with $\ell=J$, $s=0,1$ one obviously
writes $S=\eta\exp(2i\delta)$. For the states with $\ell=J \pm 1$,
$s=1$, coupled by the tensor force, six parameters are needed to
parametrize the $2\times2$ $S$ matrix. In this case it is not so
easy to think of a convenient parametrization which satisfies all
constraints from unitarity, is completely general, and free from
nontrivial ambiguities. We have used a generalization~\cite{Bry81}
of the ``bar-phase'' convention commonly used
in $N\!N$ scattering. One writes (with the notation
$\overline{\delta}_{\ell J}$ for the phase shift)
\begin{equation}
   S^J \: = \:  \exp(i\overline{\delta}) \,
                \left( \begin{array}{rr}
         \cos\overline{\varepsilon}_J & i\sin\overline{\varepsilon}_J \\
        i\sin\overline{\varepsilon}_J &  \cos\overline{\varepsilon}_J
                      \end{array} \right) \,
                        \,\, H^J \,
                \left( \begin{array}{rr}
         \cos\overline{\varepsilon}_J & i\sin\overline{\varepsilon}_J \\
        i\sin\overline{\varepsilon}_J &  \cos\overline{\varepsilon}_J
                      \end{array} \right) \,
                \exp(i\overline{\delta}) \:\: ,
\end{equation}
where $\overline{\delta}={\rm diag}(\overline{\delta}_{J-1,J},
\overline{\delta}_{J+1,J})$ and
$\overline{\varepsilon}_{J}$ is the mixing parameter.
$H^J$ is a three-parameter real and symmetric matrix representing
inelasticity. It can be diagonalized in Blatt-Biedenharn fashion
\begin{equation}
   H^J \: = \: \left( \begin{array}{rr}
                 \cos\omega_J & -\sin\omega_J \\
                 \sin\omega_J &  \cos\omega_J
                      \end{array} \right) \,
               \left( \begin{array}{cc}
                      \eta_{J-1,J} &       0        \\
                          0        &   \eta_{J+1,J}
                      \end{array} \right) \,
                \left( \begin{array}{rr}
                 \cos\omega_J & \sin\omega_J \\
                -\sin\omega_J & \cos\omega_J
                      \end{array} \right) \:\: ,
\end{equation}
where the diagonal matrix contains the ``eigeninelasticities''
$\eta_{J-1,J}$ and $\eta_{J+1,J}$, and $\omega_J$ is again a
mixing parameter. We are presently in the process of doing
a careful evaluation of these phase-shift and inelasticity
parameters and their errors. Unfortunately, this involves
a large amount of work. These and other issues will be
the subject of future publications.

\end{document}